\documentclass[11pt]{article}
\usepackage{hyperref}
\pdfoutput=1
\usepackage{hyperref}
\pdfoutput=1
\begin{document}
\title{Multiscale simulation of blood flow in brain arteries with an aneurysm}
\author{Leopold Grinberg$^1$, Vitali Morozov$^2$, Dmitry A. Fedosov$^3$, \\
        Joseph A. Insley$^2$, Michael E. Papka$^2$, Kalyan Kumaran$^2$, \\
        George Em Karniadakis$^1$
\\
\\\vspace{6pt}
$^1$ Division of Applied Mathematics, Brown University, \\ 
    Providence, RI 02906, USA
\\
\\\vspace{6pt}
$^2$ Argonne National Laboratory, \\ 
     9700 South Cass Avenue Argonne, IL 60439, USA 
\\
\\\vspace{6pt}
$^3$ Institute of Complex Systems FZ Juelich, \\ 
     Juelich, 52425, Germany} 
\maketitle
\begin{abstract}
Interfacing atomistic-based with continuum-based simulation codes is now required in many multiscale physical and
biological systems.
We present the first results from coupled atomistic-continuum simulations on 190,000 processors.
Platelet aggregation in the patient-specific model of an aneurysm has been modeled using
a high-order spectral/hp element Navier-Stokes 
solver with a stochastic (coarse-grained) Molecular
Dynamics solver based on Dissipative Particle Dynamics
(DPD).
\end{abstract}
\section{Introduction}

Low resolution movie: \href{http://www.dam.brown.edu/people/lgrinb/PRESENTATIONS/APSpackage/anc/video_entry_V013_small.mpg}{Video 1}

High resolution movie: \href{http://www.dam.brown.edu/people/lgrinb/PRESENTATIONS/APSpackage/anc/video_entry_V013_large.mpg}{Video 2}

The fluid dynamics video (in low resolution Video 1 and in high resolution Video 2) shows:
\begin{enumerate}
\item
{\bf Dissipative Particle Dynamics (DPD) simulation of healthy (red) and
malaria-infected (blue) red blood cells (RBC)}. 
The membrane of each RBC is represented by 500 DPD particles [1].
The blood plasma is also represented by DPD particles, shown as spheres colored 
by the local velocity magnitude. 
The ensemble average velocity is computed using window proper orthogonal decomposition, 
and is presented at several slices along the computational domain.
\item
{\bf Patient-specific domain of major brain arteries including circle of Willis,
internal carotid, vertebral and communicating  arteries; as well as large aneurysm}. 
At the four inlets patient-specific inflow boundary conditions
have been imposed. At the multiple outlets the pressure is regulated by the 
RC boundary conditions [2].  
The main flow characteristics are $Re=394$ and $Ws=3.7$.
Computational domain is subdivided into four overlapping patches composed of
about $0.5M$ spectral elements. In each element the solution is approximated 
with six-order polynomial expansion. 
Instantaneous velocity field is represented by streamlines, which also highlight 
the presense of an unsteady vortex inside the aneurysm. 
\item
{\bf Platelet aggregation at the wall of the aneurysm}. Simulation is performed using 
a coupled continuum-atomistic solver. The continuum solution in the macroscopic domain is obtained 
by Nektar, while the atomistic solution inside 4 $mm^3$ microscopic subdomain 
is computed by DPD-LAMMPS. The velocity boundary conditions for the microscopic 
domain are updated every 200 time-steps using data simultaneousely computed by Nektar. 
Platelet aggregation simulations have been performed on up to 190,740 processors of CRAY XT5 at ORNL
and on up to 131,072 processors of BlueGene/P at ANL.
Platelet aggregation modeled is described in [3].
More information on the coupled continuum-atomistic solver will be provided in the 
SC'11 Gordon Bell Prize finalist paper [4]. 
\end{enumerate}

[1] D. A. Fedosov, B. Caswell and G. E. Karniadakis. {\it A
multiscale red blood cell model with accurate
mechanics, rheology, and dynamics}. Biophysical
Journal, 98(10):2215–2225, 2010.

[2] L. Grinberg and G. Em Karniadakis.
{\it Outflow Boundary Conditions for Arterial Networks with Multiple Outlets}.
Annals of Biomedical Engineering, 36(9):1496-1514, 2008. 

[3] I. V. Pivkin, P. D. Richardson and G. E. Karniadakis.
{\it Blood flow velocity effects and role of activation delay
time on growth and form of platelet thrombi}.
Proceedings of the National Academy of Sciences USA,
103(46):17164–17169, 2006.

[4] L. Grinberg, V. Morozov, D. A. Fedosov, J. A. Insley, 
M. E. Papka, K. Kumaran and G. Em Karniadakis.
{\it A new Computational Paradigm in Multiscale Simulations: 
Application to Brain Blood Flow}.
In Proceedings of the
2011 ACM/IEEE International Conference for High
Performance Computing, Networking, Storage and
Analysis, SC'11.   

\end{document}